\documentclass[10pt]{article}
\usepackage{amsfonts}
\usepackage{epsfig}
\usepackage{epsf}

\def\ubar{\bar{u}}

\newcommand{\fslash}[1]{\mbox{$\!\not\!#1$}}

\begin{document}
\begin{center}
{\Large\bf Gauge Invariant Two-Photon-Exchange Contributions in $e^-\pi^+ \rightarrow e^-\pi^+$ }\\
\vspace*{1cm}
Hai. Qing. Zhou  \footnote{E-mail: zhouhq@mail.ihep.ac.cn} \\
\vspace{0.3cm} %
{\  Department of Physics, Southeast University,  Nanjing,\ 211189,\ P. R. China}\\

\vspace*{1cm}
\end{center}
\begin{abstract}
The gauge invariant two-photon exchange (TPE) contributions in $e^-\pi^+ \rightarrow
e^-\pi^+$ are discussed at hadronic level. The contact term is added to keep the full amplitude gauge invariant by two methods: one is to multiply  form factors with the amplitude for point-like particles and another is to construct a gauge invariant Lagrangian. The practical calculations show the TPE contributions by these two methods are almost the same, while the later method is favored when extending the discussion to processes including two charged finite-size particles like $ep \rightarrow en\pi^+$.

\end{abstract}

\section{Introduction}

It has been shown the two-photon exchange (TPE) contributions in unpolarized elastic $ep$ scattering
play an important role in extracting the electromagnetic form factors of the proton from the angle dependence of cross section. It is natural to expect that similar
effects may exist in the unpolarized $ep \rightarrow en\pi^+$, which is also used to
extract the electromagnetic $\pi$ form factor or $\sigma_L$ from the angle dependence of
cross section \cite{Ex-pion}. In the literature, many model dependent
calculations \cite{TPE-ep-Model-D} and model independent analyses \cite{TPE-ep-Model-UnD} have been made to
study TPE contributions in elastic $ep$ scattering, while the TPE contributions in $ep \rightarrow en\pi^+$ are much more complex and the discussion on such TPE contributions is deficient. Formally, how to
keep gauge invariance in hadronic level for such processes \cite{gauge-invariance} is a non-trivial problem, since
two finite-size charged particles play their roles.
Before discussing the gauge invariant TPE contributions in $ep \rightarrow en\pi^+$, it is a good basis to study
the gauge invariant TPE contributions in $e^-\pi^+\rightarrow e^-\pi^+$. The TPE contributions in the latter process have been studied
in \cite{Blunden2010,Dong2010}, while the contact term is usually neglected. This leads to manifest breakdown of gauge invariance. For the processes with charged non-point-like particles, the usual way to keep the full amplitude gauge invariant is to multiply form factors with the amplitudes for  point-like particles. In this letter, we introduce a gauge invariant Lagrangian to treat the TPE contributions in $e^-\pi^+\rightarrow e^-\pi^+$. Such a method can also be applied to treat the TPE contributions in $ep\rightarrow en\pi^+$ directly. We arrange our discussion as follows: in section 2, the TPE contributions in the literature are reviewed and the way to restore gauge invariance at the amplitude level is discussed; in section 3, a simple gauge invariant Lagrangian is constructed to describe the electromagnetic interactions of $\pi$ and the TPE contributions are discussed by this Lagrangian; in section 4, the numerical results are presented.

\section{Gauge Invariant TPE Contributions in $e^-\pi^+\rightarrow e^-\pi^+$:  A}

For a charged point-like  pseudoscalar particle, the electromagnetic
interaction to the lowest order can be described as
\begin{eqnarray}
L_0 = (D_\mu \phi)^* D^\mu
\phi-\frac{1}{4}F_{\mu\nu}F^{\mu\nu} \label{L_point_charge},
\end{eqnarray}
with $F_{\mu\nu}=\partial_\mu A_\nu -
\partial_\nu A_\mu$, $D_\mu=\partial_\mu+ie_QA_\mu$ and $e_Q$ being the charge of pseudoscalar particle.
To keep the gauge invariance, a contact term may be introduced by the minimal coupling. This is
different with point-like spin-$\frac{1}{2}$ particle where contact term is not necessary.

For finite-size charged pseudoscalar particles such as $\pi^+$ with $e_Q=-e=|e|$, higher order interactions are needed to
described its electromagnetic structure. In the literature \cite{Blunden2010,Dong2010}, to describe such structure a form
factor is directly multiplied with the point-like particle vertex. This
corresponds to the following replacement for the  vertex:
\begin{eqnarray}
ie(p_1+p_2)_\mu \rightarrow ie(p_1+p_2)_\mu F_\pi(q^2),
\end{eqnarray}
with $p_1,p_2,q\equiv p_2-p_1$ the momentum of incoming $\pi^+$,out coming $\pi^+$ and photon. By such
replacement, it is easy to check that the sum of amplitudes corresponding to TPE diagrams Fig.\ref{Fig:TPE-diagrams}(a) and Fig.\ref{Fig:TPE-diagrams}(b)
is not gauge invariant. To restore the gauge invariance, the contact term should be considered
as the point-like particle case, with the simple replacement is(we named it method A):
\begin{figure}[t]
\centerline{\epsfxsize 2.0 truein\epsfbox{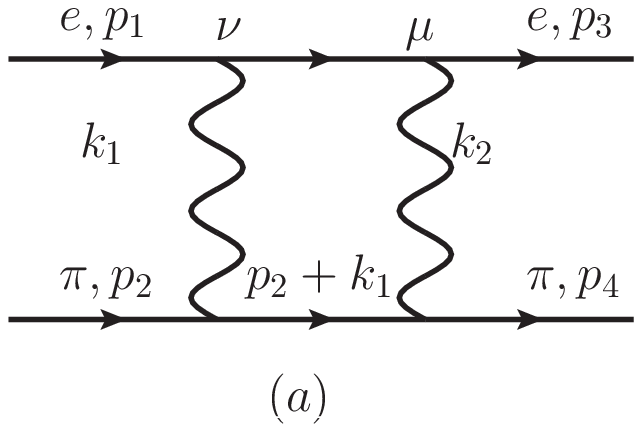}\epsfxsize 1.98
truein\epsfbox{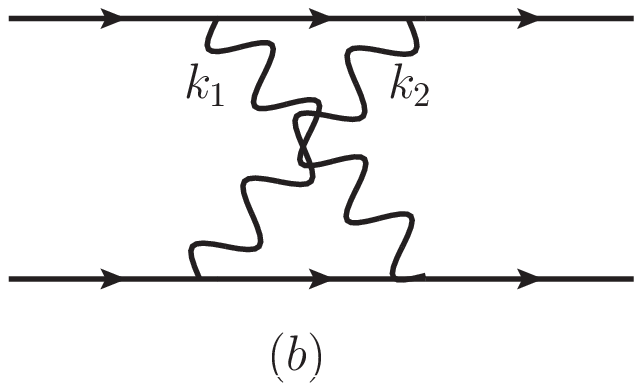} \epsfxsize 2.0 truein\epsfbox{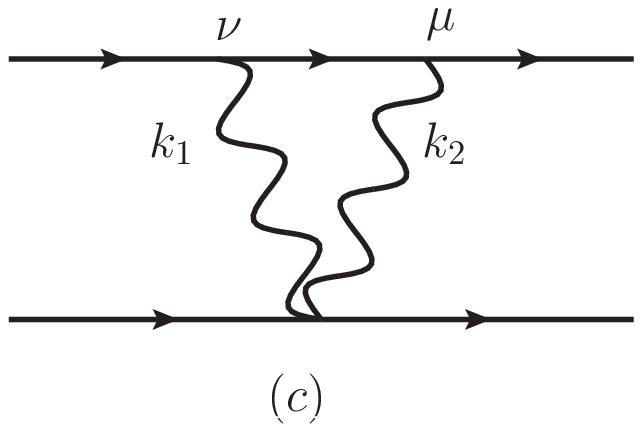} } \caption{Two-photon-exchange
diagrams with elastic intermediate state: (a) box diagram, (b) cross-box diagram and (c) contact term diagram.}
\label{Fig:TPE-diagrams}
\end{figure}

\begin{eqnarray}
 i2e^2g_{\mu\nu}
\rightarrow i2e^2g_{\mu\nu} F_\pi(k_1^2)F_\pi(k_2^2),
\end{eqnarray}
with $k_1,k_2$ the momentum of (incoming) photons and the diagram Fig.\ref{Fig:TPE-diagrams}(c) due to contact term is included.
The amplitudes corresponding to the three diagrams in Fig.\ref{Fig:TPE-diagrams} in Feynman gauge by this method read as
\begin{eqnarray}
{\cal M}_{\gamma\gamma}^{A,(a)}
&=& -i \int {d^4 k_1\over (2\pi)^4}
{\ubar_e(p_3)
 (-ie\gamma_\mu)
       ({\fslash{p_1}-\fslash{k_1} + m_e })
       (-ie\gamma_\nu) u_e(p_1)
  \over \left[(p_1-k_1)^2 - m_e^2+i\epsilon\right]\left[(p_2+k_1)^2-m_\pi^2+i\epsilon\right]}\nonumber\\
& &~~~~~~~~~~~~~{\left[ieF_\pi(k_2^2)(2 p_4-k_2)^\mu\right]\left[ieF_\pi(k_1^2)(2 p_2+k_1)^\nu \right]\over (k_1^2+i\epsilon)(k_2^2+i\epsilon)},\nonumber\\
{\cal M}_{\gamma\gamma}^{A,(b)}
&=& -i \int {d^4 k_1\over (2\pi)^4}
{\ubar_e(p_3)
 (-ie\gamma_\mu)
       ({\fslash{p_1}-\fslash{k_1} + m_e })
       (-ie\gamma_\nu) u_e(p_1)
  \over \left[(p_1-k_1)^2 - m_e^2+i\epsilon\right]\left[(p_2+k_2)^2-m_\pi^2+i\epsilon\right]}\nonumber\\
& &~~~~~~~~~~~~~{\left[ieF_\pi(k_1^2)(2 p_4-k_1)^\nu \right]  \left[ieF_\pi(k_2^2)(2 p_2+k_2)^\mu\right]\over (k_1^2+i\epsilon)(k_2^2+i\epsilon)},\nonumber\\
{\cal M}_{\gamma\gamma}^{A,(c)}
&=& -i \int {d^4 k_1\over (2\pi)^4}
{\ubar_e(p_3)
 (-ie\gamma_\mu)
       i({\fslash{p_1}-\fslash{k_1} + m_e })
       (-ie\gamma_\nu) u_e(p_1)
  \over \left[(p_1-k_1)^2 - m_e^2+i\epsilon\right]}\nonumber\\
& &~~~~~~~~~~~~~{ (-)\left[i2e^2F_\pi(k_2^2)F_\pi(k_1^2)g^{\mu\nu}\right]\over (k_1^2+i\epsilon)(k_2^2+i\epsilon)},
\end{eqnarray}
where ${\cal M}_{\gamma\gamma}^{A,(a)}$ and ${\cal M}_{\gamma\gamma}^{A,(b)}$ are the same as \cite{Blunden2010}. Now it is easy to check the full amplitude is not dependent on the gauge parameter in the photon's propagators.

With such a method based on the amplitudes directly, in principle that is not the case. And it is also not easy to extend it to processes with two
finite-size charged particles in a unitary way. In the following, we construct a simple gauge invariant Lagrangian to discuss the TPE contributions.

\section{Gauge Invariant TPE Contributions in $e^-\pi^+\rightarrow e^-\pi^+$: B}

Differently from using direct replacements as above, higher order terms can be added to describe the structure formally, one simple form being
\begin{eqnarray}
L = L_0+L_1, \label{Eq:Lagrangian}
\end{eqnarray}
with
\begin{eqnarray}
L_1 &=&ie_Q D_\mu \phi^*\phi \partial_\nu f(-\partial_\rho\partial^\rho)F^{\mu\nu}+h.c.\nonumber
\end{eqnarray}
Based on this Lagrangian the electromagnetic form factors of $\pi$ at tree level can be written as
\begin{equation}
\begin{array}{lll}
<p_2|J_\mu|p_1>=(1+q^2f(q^2))(p_1+p_2)_\mu.
\end{array}
\end{equation}
Comparing with the general form of electromagnetic form factor of the $\pi$
\begin{equation}
\begin{array}{lll}
<p_2|J_\mu|p_1>=F_\pi(q^2)(p_1+p_2)_\mu,
\end{array}
\end{equation}
the following relation is obtained
\begin{equation}
\begin{array}{lll}
F_\pi(q^2)&=&1+q^2f(q^2).\\
\label{relation}
\end{array}
\end{equation}
In principle, the Lagrangian Eq.(\ref{Eq:Lagrangian}) is not the most general one,
while it is the simplest one to keep the gauge invariance in a
manifest way. With Lagrangian Eq.(\ref{Eq:Lagrangian}), the amplitudes in Feynman gauge for the three diagrams in Fig.\ref{Fig:TPE-diagrams}
can be expressed as
\begin{eqnarray}
{\cal M}_{\gamma\gamma}^{B,(a)}
&=& -i \int {d^4 k_1\over (2\pi)^4}
{\ubar_e(p_3)
 (-ie\gamma_\mu)
       ({\fslash{p_1}-\fslash{k_1} + m_e })
       (-ie\gamma_\nu) u_e(p_1)
  \over \left[(p_1-k_1)^2 - m_e^2+i\epsilon\right]\left[(p_2+k_1)^2-m_\pi^2+i\epsilon\right]}\nonumber\\
  &&~~~~~~~~~~~~~{\Gamma^\mu(p_4,p_4-k_2)\Gamma^\nu(p_4-k_2,p_2)\over (k_1^2+i\epsilon)(k_2^2+i\epsilon)},\nonumber\\
{\cal M}_{\gamma\gamma}^{B,(b)}
&=& -i \int {d^4 k_1\over (2\pi)^4}
{\ubar_e(p_3)
 (-ie\gamma_\mu)
       ({\fslash{p_1}-\fslash{k_1} + m_e })
       (-ie\gamma_\nu) u_e(p_1)
  \over \left[(p_1-k_1)^2 - m_e^2+i\epsilon\right]\left[(p_2+k_2)^2-m_\pi^2+i\epsilon\right]}\nonumber\\
  &&~~~~~~~~~~~~~{\Gamma^\mu(p_4,p_4-k_1)\Gamma^\nu(p_4-k_1,p_2)\over (k_1^2+i\epsilon)(k_2^2+i\epsilon)},\nonumber\\
{\cal M}_{\gamma\gamma}^{B,(c)}
&=& -i \int {d^4 k_1\over (2\pi)^4}
{\ubar_e(p_3)
 (-ie\gamma_\mu)
       i({\fslash{p_1}-\fslash{k_1} + m_e })
       (-ie\gamma_\nu) u_e(p_1) (-)\Lambda^{\mu\nu}(k_1,k_2)
  \over \left[(p_1-k_1)^2 - m_e^2+i\epsilon\right](k_1^2+i\epsilon)(k_2^2+i\epsilon)},
\end{eqnarray}
with
\begin{eqnarray}
\Gamma^\mu(p_f,p_i) &=& ie\left[(1+f(q^2)q^2)(p_f+p_i)^\mu-f(q^2)(p_f^2-p_i^2)q^\mu \right],\nonumber\\
\Lambda^{\mu\nu}(k_1,k_2)&=& 2ie^2 \left[g^{\mu\nu}+f(k_1^2)(k_1^2g^{\mu\nu}-k_1^\mu k_1^\nu)+f(k_2^2)(k_2^2g^{\mu\nu}-k_2^\mu k_2^\nu) \right].
\end{eqnarray}

\section{ Results}
To show the TPE contributions,we define
\begin{eqnarray}
\delta_{(a)/(b)/(c)}^{A/B}={ 2 Re \{ {\cal M}_0^*\, {\cal M}_{\gamma\gamma}^{A/B,(a)/(b)/(c)} \}
           \over |{\cal M}_0|^2}\, ,
\end{eqnarray}
with ${\cal M}_0$ the one photon exchange amplitude, A/B refer to Method A/B and (a)/(b)/(c) refer to corresponding diagrams, respectively.
In Feynman gauge, we can prove the sum  $\delta_{(a)+(b)}^A$ is equal to $\delta_{(a)+(b)}^B$ with any form factors as input though  $\delta^A_{(a)/(b)}$  are not equal to $\delta^B_{(a)/(b)}$, respectively. Generally such equivalence is not true for other gauge parameters. And the contributions from diagrams $(c)$ are not equivalent by the two methods.

To show the detail, we take the same form of $F_\pi(q^2)$ with \cite{Blunden2010}
\begin{eqnarray}
F_\pi(q^2)=\frac{-\Lambda^2}{q^2-\Lambda^2},
\end{eqnarray}
with $\Lambda=0.77$GeV.

\begin{figure}[t]
\centerline{\epsfxsize 6.0 truein\epsfbox{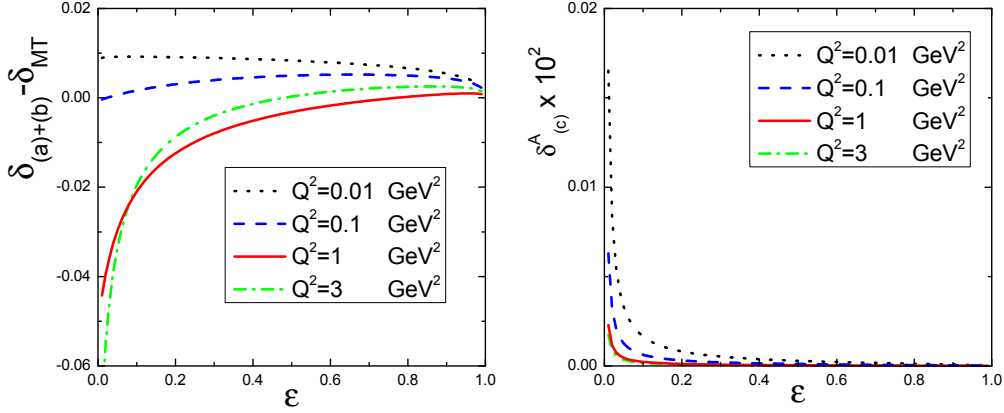}} \caption{Two-photon-exchange
contributions: the left panel is for $\delta_{(a)+(b)}-\delta_{MT}$ vs. $\varepsilon$ and the right panel is for $\delta^{A}_{(c)}\times10^2$ vs. $\varepsilon$ both with $Q^2=0.01,0.1,1,3$GeV$^2$.} \label{Fig-delta-epslion}
\end{figure}

With this monopole form factor as input the TPE contributions can be calculated directly. And we subtract the IR divergence in the
same way as \cite{Blunden2010}. The left panel of Fig.\ref{Fig-delta-epslion} shows $\delta_{(a)+(b)}-\delta_{MT}$($\equiv \delta_{(a)+(b)}^{A/B}-\delta_{MT}$) vs. $\varepsilon$ in Feynman gauge where
$\varepsilon = \left(1 + 2 (1+\tau) \tan^2{(\theta/2)}\right)^{-1},\tau = Q^2/4m_\pi^2,Q^2=-(p_4-p_2)^2$, $\theta$ the scattering angle and $\delta_{MT}$
denotes the correction from the box diagrams in the soft photon approximation given by the standard treatment of Mo and Tsai \cite{MoTsai}. The right panel of Fig.\ref{Fig-delta-epslion} shows $\delta_{(c)}^A$ vs. $\varepsilon$. The practical calculation shows the corrections $\delta_{(c)}^{A/B}$ in Feynman gauge are about $10^{-5}\sim10^{-6}$ in almost all $\varepsilon$ region by both two methods for $Q^2$ from $0.01$GeV$^2$ to $3$GeV$^2$. The relative magnitudes $\delta^{A}_{(c)}/\delta^{B}_{(c)}$ are shown in Tab.\ref{Tab:1}. An interesting property is that $\delta^{A}_{(c)}/\delta^{B}_{(c)}$ are independent on $\varepsilon$. They are very small when $Q^2<1$GeV$^2$ and increase with $Q^2$. The small $\delta_{(c)}^{A/B}$ result in almost the  same full TPE contributions by the two methods. This means the main results by \cite{Blunden2010} are kept, while this does not mean the contact term can be neglected in other processes or other gauges. When extending the calculation to $ep \rightarrow en\pi^+$, the contributions from such term need to be considered more carefully and the method B is favored because of the manifest gauge invariance.

\begin{table}[htbp]
\vspace{0.5cm}
\begin{tabular}
{|p{60pt}|p{70pt}|p{70pt}|p{70pt}|p{70pt}|}
\hline     &$Q^2=0.01$GeV$^2$ &$Q^2=0.1$GeV$^2$ &$Q^2=1$GeV$^2$ &$Q^2=0.3$GeV$^2$ \\
\hline $\delta^{A}_{(c)}/\delta^{B}_{(c)}$ & 1.0002& 1.0014& 1.0103& 1.0280 \\
\hline
\end{tabular}
\caption{Numerical results for $\delta^{A}_{(c)}/\delta^{B}_{(c)}$ with $Q^2=0.01,0.1,1,3$GeV$^2$.}
\label{Tab:1}
\end{table}

\section{Acknowledgment}
This work is supported by the National Sciences Foundations of China
under Grant No.10805009. The author gladly acknowledges the
support of Theoretical Physics Center for Science Facilities for his visit where part of this work is finished.


\end{document}